\documentclass[11pt]{article}

\usepackage{amsmath}
\usepackage{amssymb}
\usepackage{geometry}
\usepackage{graphicx}
\usepackage{xurl}
\usepackage{algorithm}
\usepackage{tikz}
\usepackage{xcolor}
\usepackage{natbib}
\usepackage{float}
\usetikzlibrary{shapes.geometric, arrows}
\usepackage[hidelinks]{hyperref}
\usepackage{algpseudocode}

\title{Surviving the Unseen: Predictive Defense for Novel Multi-Turn Multimodal Attacks}
\author{
    Doohee You\thanks{Corresponding author: \texttt{doohee@google.com} The views and opinions expressed in this paper are solely those of the authors and do not necessarily reflect the official policy, position, or views of their respective employers.} \\
    \textit{Trust and Safety, Google} \\ 
}

\date{}
\date{}

\begin{document}

\maketitle

\begin{abstract}
The expansion of Multimodal Large Language Models (MLLMs) and their integration into autonomous agentic workflows has introduced a non-stationary attack surface. Empirical observations indicate that adversaries employ progressive, cross-modal perturbations that evade turn-specific guardrails by distributing malicious intent across longitudinal conversational trajectories. Static defense mechanisms, constrained by the Markov property, evaluate inputs in isolation and fail to detect cumulative structural poisoning. To handle this limitation, this paper formulates safety verification as a dynamic survival prediction and trajectory dynamics problem. The Triple-tier Anomaly Defense (TRIAD) framework is proposed as a predictive model that maps multimodal and multi-turn conversational flow as a continuous trajectory. The framework integrates structural anomaly detection to monitor covariance shifts, a Ledoit-Wolf regularized Mahalanobis distance to monitor covariance shifts in high-dimensional spaces, and topological trajectory acceleration to differentiate benign creative exploration from continuous malicious drift. These kinematic and geometric features are integrated into a time-varying Cox Proportional Hazards model via a Bayesian Hidden Markov Model (HMM) feedback loop. Theoretical analysis demonstrates that the TRIAD framework provides a mathematically bounded expected time-to-failure under adversarial perturbations, ensuring that malicious acceleration diverges positively. This framework provides a computationally efficient, interpretable, and predictive safeguard for real-time agentic AI systems, establishing a rigorous foundation for continuous safety alignment without relying on empirical retraining.
\end{abstract}

\section{Introduction}

Generative Artificial Intelligence is transitioning from single-turn text  systems into models capable of processing visual, auditory, and text information simultaneously over multi-turn interactions \cite{openai2023gpt4, anil2023gemini, qi2024visual}. While the integration of multi-modality maximizes reasoning capabilities and operational versatility, it introduces security vulnerabilities that are unseen in unimodal text environments where risk is measured via static snapshots. These vulnerabilities reveal a blind spot in legacy safety alignment systems, which often prioritize task instruction following over pre-programmed safety constraints \citep{bailey2023image}. Unlike small-scale applications where escalated issues can be manually reviewed case-by-case, production environments facing a massive user base require swift detection of unseen, nuanced adversarial attack patterns to prevent the unintentional generation of harmful content. Tackling these issues is a critical defense mechanism for maintaining continuous safeguard alignment.

When adversaries induce attention mismatches between text and image modalities, or progressively inject subtle instructions that appear harmless in individual turns,  MLLMs showed their weakness to maintain their aligned safety policies \cite{wei2023jailbroken, zou2023universal}. This phenomenon functions as a cross-modal alignment tax. Static text-based guardrails and single-point binary classifiers deployed in production environments are constrained by the Markov Property; they evaluate the information of a specific input turn independently, lacking historical context \citep{dong2026robust}. Consequently, these defenses exhibit limitations in detecting structural poisoning caused by the long-term accumulation of subtle malicious contexts. If adversarial perturbations are distributed below the detection threshold of individual turns, the system ultimately loses control without recognizing the compounding risk \citep{russinovich2024crescendo, liu2024autodan}.

The current empirical paradigm that relies heavily on large-scale defense datasets and supervised learning against specific, known attack signatures, this study investigated generative AI response in multimodal and multi-turn environment as a dynamic survival prediction and trajectory dynamics problem over time. By treating the internal latent space of model as a multi-dimensional state space and modeling the continuous flow of conversation as a topological trajectory, a hybrid statistical framework is established to detect unknown zero-day attacks. The proposed defense mechanism is structured into a continuous analytical pipeline. 1) First, we employ Isolation Forest \cite{liu2008isolation} to identify structural isolation within the high-dimensional embedding space. 2) Second, it quantifies distributional drift via a robust distance metric that resolves singular matrix issues common in high-dimensional multimodal spaces \cite{ledoit2004well}. 3) Third, it calculates trajectory acceleration via the second derivative of the phase space distance to statistically separate human creative exploration from adversarial evasion. 4) Finally, it predicts the time-to-failure hazard by integrating time-series indicators into a Survival Analysis model \citep{cox1972regression}. This research provides mathematical validity that the proposed framework yields a bounded detection rate under unknown cross-modal poisoning. By adopting a computationally efficient anomaly detection approach, we ensure real-time feasibility for light immediate application in real-time agentic systems.

\section{Related Work}

Research on prompt jailbreaks and corresponding defenses in single-modality environments has been extensively conducted, but the advent of large-scale Vision-Language Models (VLMs) has expanded the attack surface \cite{qi2024visual, bailey2023image}. Optimized visual adversarial examples can neutralize the guardrails of safety-aligned language models, demonstrating that continuous, high-dimensional visual inputs bypass text-token-based filtering \cite{wei2023jailbroken}. The security vulnerabilities of MLLMs under multi-turn prompting have become a central focus. Recent literature confirms vulnerabilities against automated multi-turn jailbreaks, such as Crescendo, where malicious intent is fragmented across interactions \cite{russinovich2024crescendo, liu2024autodan}. Also, recent frameworks like PolyJailbreak demonstrate that visual alignment introduces uneven safety constraints across modalities, leading to a phenomenon identified as multimodal safety asymmetry \cite{wang2026polyjailbreak}. To exploit this, Foot-In-The-Door (FITD) methods leverage psychological manipulation to bypass safeguards across multiple turns by establishing benign context before escalating requests \cite{weng2025foot}. Advanced automated systems such as Mastermind employ a hierarchical multi-agent architecture with strategy-space fuzzing to autonomously discover and refine these multi-turn jailbreaks \cite{li2026knowledge}. To counter these threats, the Multi-Turn Safety Alignment (MTSA) framework uses thought-guided attack learning and multi-turn reinforcement learning to improve the safety alignment of target models \cite{guo2025mtsa}.

As Agentic AI systems transition to production, the threat landscape shifts toward indirect prompt injection, goal hijacking, and tool misuse. Unlike standard LLMs in read-only sandbox environments, Agentic AI systems possess read-write API access and persistent storage, escalating the impact of breaches from session-based misinformation to systemic compromise \cite{garg2023multimodal}. Evaluative frameworks confirm that existential and systemic safety remains a structural weakness, with current safety practices demonstrating partial alignment with emerging global standards but lacking rigorous implementation. Defensive strategies are shifting toward dynamic agent red teaming and test-time immunization, which aims to defend AI systems at the inference stage without requiring complete model retraining. To defend these multi-turn multimodal jailbreaks directly, systems use fragment-optimized MLLM defense mechanisms to systematically analyze and mitigate progressive attacks without requiring fine-tuning \cite{das2026multiturn}.

Out-of-Distribution (OOD) detection remains a core area of AI security for identifying unknown data patterns \cite{lee2018simple, ren2019likelihood}. Previous frameworks utilized Gaussian Discriminant Analysis and Mahalanobis distance; however, directly applying this to the latent space of multimodal models encounters mathematical hurdles. To address the rank-deficiency and numerical instability inherent in high-dimensional multimodal manifolds\cite{wang2022high}, we move beyond vanilla Mahalanobis metrics by incorporating robust shrinkage-based covariance estimation \cite{ledoit2004well}. Recent advancements attempt to leverage the multimodal reasoning capabilities of MLLMs, synthesizing pseudo-OOD data from convex combinations of in-distribution data to enhance discrimination. Despite these advancements, robust statistical estimation integrating shrinkage techniques is required to mathematically handle the high dimensionality of multimodal embeddings.

To bypass strict Gaussian assumptions altogether, anomaly detection methodologies focusing on topological structure are utilized. While frameworks like Isolation Forest are traditionally deployed for large-scale network logs and API abuse detection, adapting them to monitor continuous embedding trajectories within multimodal interactions represents a paradigm shift \citep{liu2008isolation, pang2021deep, zhao2023detecting}. Recent paradigms build on this by combining embedding-based anomaly detection with LLM-driven validation frameworks. These systems dynamically address diverse anomaly types by pairing dimensionality reduction with Isolation Forests to filter out false positives typical of traditional machine learning detectors.

Rather than treating multi-turn interactions as independent events, we frame adversarial detection as a longitudinal survival problem. This approach is uniquely suited to capture the cumulative hazard of subtle semantic drifts—risks that are often invisible to memoryless, snapshot-based guardrails \cite{cox1972regression, katzman2018deepsurv}. The application of survival analysis to LLM alignment has emerged as a method to model the robustness of models in extended multi-turn dialogues. Models such as Time-To-Inconsistency evaluate conversational failure via survival analysis, showing that gradual semantic drift is protective, while abrupt, prompt-to-prompt drift increases the hazard of policy violations \cite{li2025time}. These studies analyze thousands of conversation turns to demonstrate that abrupt, prompt-to-prompt semantic drift increases the hazard of conversational failure. AFT models with model-drift interactions have achieved high discrimination and calibration in these contexts. However, these approaches rely heavily on semantic surface indicators within a single text modality, highlighting the need to apply survival analysis to the kinematic properties of multimodal latent spaces.

\section{Problem Formulation and Framework Architecture}

To overcome the limitations of static evaluation, the multimodal multi-turn interaction between a user and a Generative AI model is modeled as continuous state changes within a high-dimensional phase space. The interaction time-step is defined as discrete time $t \in \mathbb{N} = \{1, 2,..., T\}$. At turn $t$, the user provides a text prompt $X_{\text{text}}^{(t)}$ and image data $X_{\text{img}}^{(t)}$, which the corresponding encoders map to latent representations $E_{\text{text}}^{(t)} \in \mathbb{R}^{d_{\text{text}}}$ and $E_{\text{img}}^{(t)} \in \mathbb{R}^{d_{\text{img}}}$. An integrated cross-modal semantic vector is generated using a concatenation operator:
\begin{equation}
V_{semantic}^{(t)} = E_{text}^{(t)} \oplus E_{img}^{(t)} \in \mathbb{R}^{d_{\text{semantic}}}.
\end{equation}

To statistically separate the mechanical characteristics of an automated attacker from the cognitive reflection of a normal user, the Contextual Covariate Modulator (CCM) introduces a behavioral covariates vector $B^{(t)} \in \mathbb{R}^{d_{behav}}$. While known classical behavioral biometrics exploit static typing anomalies to detect localized automation \cite{mehta2026detectingllmassistedacademicdishonesty}, our framework treats these artifacts as a subset of a broader thermodynamic phenomenon: the low-entropy boundary condition of automated scripts. Because any automated adversarial framework—regardless of its underlying prompt architecture or unseen optimization target—must algorithmically dispatch tokens, it inherently collapses the temporal and physical variance of the interaction manifold. Conversely, benign creative exploration is characterized by high-entropy behavioral distributions, manifesting as stochastic temporal variations, irregular cognitive pauses, and non-linear trajectory accelerations across the interaction manifold. The final input state vector is defined as the concatenation of semantic and behavioral data:
\begin{equation}
V^{(t)} = V_{semantic}^{(t)} \oplus B^{(t)} \in \mathbb{R}^{D},
\end{equation}
where $D = d_{\text{semantic}} + d_{\text{behav}}$.

Traditional safety models evaluate a single vector $V^{(t)}$ to determine binary safety. However, under advanced multi-turn attacks, the maliciousness of initial turns approaches zero. The random variable $T$ is defined as the number of turns at which the model first violates the aligned safety policy. The framework tracks the Survival Function $S(t) = \mathbb{P}(T > t)$, the probability that the dialogue survives without safety violations up to turn $t$. Correspondingly, the instantaneous rate at which the model loses control at the next turn $t$ is formulated as the Hazard Function $h(t)$:
\begin{equation}
h(t) = \lim_{\Delta t \rightarrow 0} \frac{\mathbb{P}(t \le T < t + \Delta t \mid T \ge t)}{\Delta t}.
\end{equation}
If malicious multi-turn perturbations are progressively injected, $S(t)$ drops exponentially alongside an increase in the hazard function $h(t)$.

To minimize latency while maximizing detection capability for real-time services, the TRIAD framework adopts a trigger-based cascade architecture combined with a Bayesian feedback loop.


\begin{algorithm}[htbp]
\caption{Dynamic State-Space Inference Pipeline}
\label{alg:pipeline}
\begin{algorithmic}[1]
\Require Sequence $\{X_{\text{text}}^{(t)}, X_{\text{img}}^{(t)}\}$, Behavioral data $B^{(t)}$, Params $\mu, \hat{\Sigma}_{\text{LW}}^{-1}$, thresholds $\tau_{\text{hazard}}, \alpha$
\State Initialize prior $\mathbb{P}(S_0) \leftarrow \text{Safe}$
\For{each turn $t = 1, 2, \dots$}
    \State $V^{(t)} \leftarrow \text{Fusion}(X_{\text{text}}^{(t)}, X_{\text{img}}^{(t)}) \oplus B^{(t)}$
    \State $S_{\text{iso}}^{(t)} \leftarrow \text{iForest}(V^{(t)})$
    \If{$S_{\text{iso}}^{(t)} > \alpha$}
        \State $D_M^{(t)} \leftarrow \sqrt{(V^{(t)} - \mu)^T \hat{\Sigma}_{\text{LW}}^{-1} (V^{(t)} - \mu)}$
        \State $a_t \leftarrow D_M^{(t)} - 2D_M^{(t-1)} + D_M^{(t-2)}$
        \State Update HMM Belief $\mathbb{P}(S_t \mid V^{(1:t)})$ using $D_M^{(t)}$ and $a_t$
        \State $h(t) \leftarrow h_0(t) \exp(\beta_1 D_M^{(t)} + \beta_2 S_{\text{iso}}^{(t)} + \gamma a_t)$
        \If{$h(t) > \tau_{\text{hazard}}$ \textbf{and} $a_t > 0$}
            \State \textbf{return} \text{ALERT: Imminent failure. Early-stop Session.}
        \EndIf
    \EndIf
\EndFor
\State \textbf{return} \text{SAFE: Continue interaction.}
\end{algorithmic}
\end{algorithm}

Performing expensive matrix inversions across the entire phase space for every turn $t$ is computationally inefficient. An Isolation Forest (iForest) ensemble is deployed as a primary unsupervised scout, possessing a low time complexity of $\mathcal{O}(\log N)$. Given the expected path length $\mathbb{E}[h(V^{(t)})]$ required to isolate a vector through random feature space partitioning, where the expected path length in a binary search tree is defined by $c(n) = 2H(n-1) - \frac{2(n-1)}{n}$ and $H(i)$ is the harmonic number, the Anomaly Score $S_{iso}(V^{(t)})$ is calculated:
\begin{equation}
S_{iso}(V^{(t)}) = 2^{-\frac{\mathbb{E}[h(V^{(t)})]}{c(n)}}.
\end{equation}
This score serves as a conditional trigger. If $S_{iso}(V^{(t)}) > \alpha$, the system allocates resources to execute precision computations.

\begin{figure}[h]
\centering
\begin{tikzpicture}[node distance=2.2cm, auto,
    box/.style={rectangle, draw, fill=blue!10, text width=8cm, text centered, rounded corners, minimum height=1.2cm},
    decision/.style={diamond, draw, fill=green!10, text width=2.5cm, text centered, inner sep=0pt},
    arrow/.style={->, thick}]

    \node[box] (input) {Multimodal Input \& \\
    Telemetric Covariates $V^{(t)}$};
    \node[box, below of=input] (iforest) {Pillar 1: Structural Scout (Isolation Forest) \\ Calculate $S_{iso}^{(t)}$};
    \node[decision, below of=iforest, node distance=2.8cm] (trigger) {$S_{iso}^{(t)} > \alpha$};
    \node[box, below of=trigger, node distance=2.8cm] (mahalanobis) {Pillar 2: Distributional Anchoring \& \\
    Kinematics \\ Calculate $D_M^{(t)}$ and $a_t$};
    \node[box, below of=mahalanobis] (hmm) {CCM: Bayesian Belief Update \\ HMM State Tracking};
    \node[box, below of=hmm] (survival) {Pillar 3: Survival Forecast \\ Cox Hazard $h(t)$};
    \node[box, below of=survival] (action) {Action: Early-stop \\
    (if $h(t) > \tau_{hazard}$) or Continue};

    \draw[arrow] (input) -- (iforest);
    \draw[arrow] (iforest) -- (trigger);
    \draw[arrow] (trigger) -- node[anchor=east] {Yes} (mahalanobis);
    \draw[arrow] (mahalanobis) -- (hmm);
    \draw[arrow] (hmm) -- (survival);
    \draw[arrow] (survival) -- (action);

    \draw[arrow] (trigger.east) -- ++(4.5,0) |- node[pos=0.25, anchor=west] {No (Safe)} (action.east);
\end{tikzpicture}
\caption{Operational Flow of the TRIAD Framework}
\label{fig:flowchart}
\end{figure}
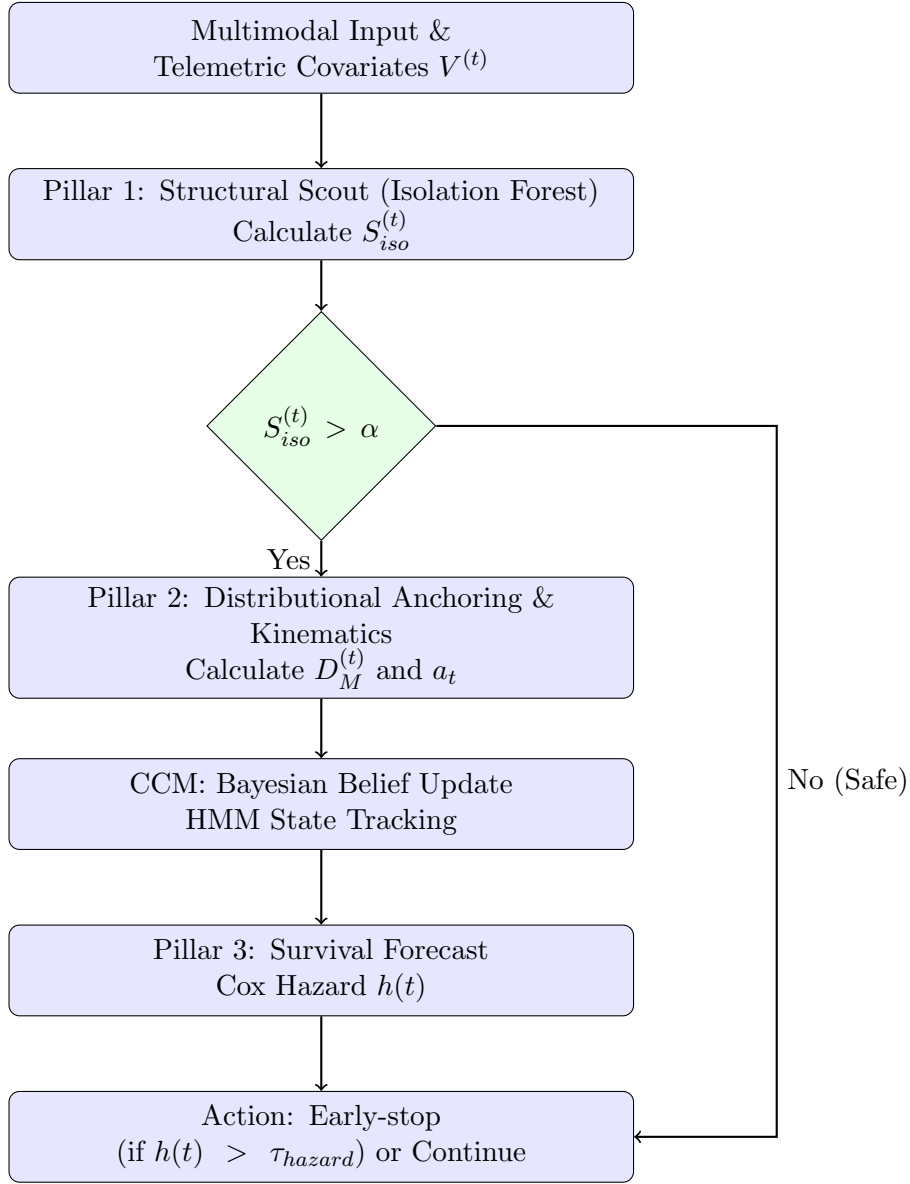

When the system is triggered, the cross-modal covariance shift is measured. In high-dimensional spaces $(p \gg n)$, calculating an accurate inverse covariance matrix $\Sigma^{-1}$ is unstable due to rank deficiency. The sample covariance matrix is defined as $\Sigma = \frac{1}{n-1}\sum_{i=1}^n (x_i - \bar{x})(x_i - \bar{x})^T$. To resolve the singular matrix problem, the Ledoit-Wolf Shrinkage estimator is employed. This method shrinks the empirical covariance matrix towards a well-conditioned target matrix, typically the identity matrix $I$, yielding:
\begin{equation}
\hat{\Sigma}_{LW} = (1 - \lambda)\Sigma + \lambda \frac{\text{Tr}(\Sigma)}{p} I ,
\end{equation}
where $\lambda$ is optimized via Frobenius norm minimization. The robust Mahalanobis Distance is subsequently calculated:
\begin{equation}
D_M(V^{(t)}) = \sqrt{(V^{(t)} - \mu)^T \hat{\Sigma}_{LW}^{-1} (V^{(t)} - \mu)}.
\end{equation}

Human dialogue naturally diverges temporarily to unfamiliar topics before recovering. Treating turns as disconnected independent events causes false permanent blocks. Within the CCM, the hidden state $S_t \in \{\text{Safe}, \text{Malicious}\}$ is mapped into a Hidden Markov Model (HMM). The posterior probability calculated in turn $t-1$ transfers as the prior probability for turn $t$, granting the system contextual inertia. This Bayesian state tracking ensures that the system evaluates current distances in the context of preceding trajectory stability.

To mathematically distinguish between a normal user's complex role-playing and an adversary's progressive attack, the topological Trajectory Acceleration is measured. Acceleration is approximated via the second-order derivative of the distance over discrete turns:
\begin{equation}
a_t = \frac{d^2}{dt^2} D_M(t) \approx D_M(t) - 2D_M(t-1) + D_M(t-2).
\end{equation}
A benign user's creative search stabilizes over time within a new local manifold, causing the acceleration to converge toward zero or become negative. Conversely, a malicious Crescendo attack must continuously twist and push the trajectory out of the manifold to bypass alignment guardrails, exhibiting an acceleration that strictly diverges in the positive direction ($a_t > 0$).

The extracted geometric isolation and kinematic acceleration indicators are processed into time-varying covariates $Z(t)$. These are injected into a Cox Proportional Hazards model to predict the instantaneous rate at which the model loses control:
\begin{equation}
h(t \mid Z(t)) = h_0(t) \exp \left( \sum_{i=1}^k \beta_i Z_i(t) \right).
\end{equation}
The partial likelihood function maximized during training is defined as:
\begin{equation}
L(\beta) = \prod_{i=1}^k \frac{\exp(\beta^T Z_i(t_i))}{\sum_{j \in R(t_i)} \exp(\beta^T Z_j(t_i))}.
\end{equation}
The term $h_0(t)$ represents the baseline hazard. Because the covariates $Z(t)$ and the positive acceleration metric $a_t$ sit in the exponent, continuous adversarial drift creates a compound multiplier effect on the marginal hazard function $h(t)$.

\section{Theoretical Analysis}

This section rigorously demonstrates how the proposed state-space framework mathematically bounds the time of model failure under unknown multi-turn attacks, derived from the mathematical properties of the statistical shape modeling structure.

\textbf{Theorem 1: Bounded Time-to-Failure under Cross-modal Poisoning.} Assume an adversary applies a subtle malicious perturbation vector to the visual latent space to generate a manipulated image embedding $\tilde{E}_{img}^{(t)} = E_{img}^{(t)} + \delta$. This perturbation possesses an $L_2$ norm $||\delta||$ sufficient to induce malicious bias during the language model's multi-head cross-attention process, while maintaining a direction geometrically orthogonal to the existing text manifold. If the perturbation is continuously injected into the joint latent space, the time-series cumulative hazard function $H(t)$ estimated by the framework possesses a monotonically increasing property. The maximum number of conversation turns $T_{max}$ the model can endure before its survival probability $S(t)$ drops below a specific safety threshold $\tau$ is mathematically bounded within a finite upper bound.

\textbf{Proof:} Let the covariance matrix calculated in the normal state's multivariate joint embedding space be represented using block matrix partitioning. Substituting the joint state vector of a new discrete time-step with the perturbation vector added as $\tilde{V}^{(t)} = V^{(t)} + [0, \delta]^T$ into the Ledoit-Wolf distance metric projection yields:
\begin{equation}
D_M(\tilde{V}^{(t)})^2 = (\tilde{V}^{(t)} - \mu)^T \hat{\Sigma}_{LW}^{-1} (\tilde{V}^{(t)} - \mu).
\end{equation}
Expanding the equation by the distributive law yields:
\begin{equation}
D_M(\tilde{V}^{(t)})^2 = D_M(V^{(t)})^2 + 2 [0, \delta]^T \hat{\Sigma}_{LW}^{-1} (V^{(t)} - \mu) + \delta_{sub}^T (\hat{\Sigma}_{LW}^{-1})_{I,I} \delta_{sub}.
\end{equation}

By the properties of matrix regularization, the Ledoit-Wolf estimate applied with the shrinkage scalar $\lambda > 0$ is guaranteed to be a strictly positive definite matrix with a stabilized condition number. By definition, for any non-zero vector ($\delta \ne 0$), the quadratic form $\delta^T \hat{\Sigma}_{LW}^{-1} \delta$ yields a strictly positive scalar. The squared distance increment induced by the perturbation $\Delta D_M(t)^2$ secures a lower bound constant $c||\delta||^2 > 0$ (where $c$ is the minimum eigenvalue $\lambda_{min}$ of the inverse matrix).

Approximating the continuous Cox survival function into a series of discrete time-steps results in:
\begin{equation}
S(t) \approx \exp \left( - \sum_{k=1}^t h_0(k) \exp(\beta_1 D_M(k) + \beta_2 S_{iso}(k)) \right).
\end{equation}
As malicious perturbations intervene at every turn, a strictly positive exponential increment is added to the covariate component. As $t \rightarrow \infty$, according to the divergence test of infinite series, the summation diverges toward infinity. Consequently, the dependent survival probability function $S(t) = \exp(-\infty)$ rapidly meets an exponential convergence to 0. It is proven that the point $T_{max}$ satisfying the minimum safe survival tolerance $\tau > 0$ must finitely exist.

\textbf{Theorem 2: Positive Divergence of Adversarial Acceleration.} Assume normal creative exploration converges to a local manifold $\mu_{local}$ representing a topological transition. Malicious Crescendo attacks follow continuous drift Brownian motion, constantly diffusing to evade detection thresholds without settling. Under continuous adversarial drift required to bypass alignment guardrails, the second-order derivative (trajectory acceleration) $a_t = \frac{d^2}{dt^2}D_M(t)$ remains strictly positive, distinguishing it mathematically from the asymptotic stabilization of benign creative shifts.

\textbf{Proof:} Let $V(t)$ be the state position function parameterized by discrete time $t$. Under a benign creative search scenario, $\lim_{t \to \infty} V(t) = \mu_{local}$ holds true. By the chain rule of calculus, the first derivative of the Mahalanobis distance scalar function $D_M(V(t))$ is derived as $\nabla D_M \cdot V'(t)$. As $V(t)$ approaches the new local settlement point, the magnitude of the velocity vector $V'(t)$ generated at each turn asymptotically decays toward 0. Hence, the trajectory acceleration function $\frac{d^2}{dt^2}D_M(t)$ converges to 0 or becomes negative.

Conversely, for an adversarial attacker to bypass the aligned filter network, they must continuously inject a forced displacement $\Delta V$ above a constant threshold at every time-step ($||V'(t)|| \ge c > 0$ for a valid constant $c$). Since such a trajectory continuously causes friction against the structural cohesion of the model's existing covariance matrix, the magnitude of the directional gradient component $\nabla D_M$ shows a monotonically increasing form. In the second derivative expansion equation:
\begin{equation}
\frac{d^2}{dt^2}D_M(t) = V'(t)^T H_{D_M} V'(t) + \nabla D_M \cdot V''(t).
\end{equation}
(where $H_{D_M}$ is the Hessian matrix), the sum of the two terms fails to turn negative due to the continuously added drift tension. Therefore, malicious acceleration never converges to 0 and always exceeds a positive lower bound.

\section{Practical Deployment and Experimental Validation Guidelines}
Deploying state-space statistical estimation models in production enterprise environments necessitates mitigating inherent structural constraints and operational edge cases. Instead of anchoring validation to volatile, model-specific empirical benchmarks, this section establishes a rigorous \textit{Experimental Protocol} and evaluation framework to systematically stress-test multi-turn agentic defenses against unknown adversarial drift.

\subsection{Architectural Mitigations for Operational Constraints}
\textbf{Manifold Heterogeneity via GMM:} The core mathematical derivation initially assumes that the multimodal interaction manifold approximates a single Multivariate Gaussian Distribution. However, in diverse production deployments, the joint latent space exhibits multi-peak distributions representing disparate topical clusters. Utilizing a single global covariance matrix causes the distance calculation to artificially expand when a user naturally switches conversational domains, inducing a high rate of false positives. 
To mitigate this, implementations should adopt a Gaussian Mixture Model (GMM) framework. During runtime, the system should probabilistically identify the local cluster $K$ most appropriate for the current semantic context, calculating the regional deviation distance using the regional mean $\mu_K$ and the regional shrunken covariance matrix $\hat{\Sigma}^{-1}_{K, LW}$.

\textbf{Temporal Inertia via AFT Models:} The Cox-based survival model is predicated on the Proportional Hazards Assumption, making it vulnerable to sudden shock attacks. An adversary could orchestrate a high-latency sequence to accumulate safe inertia within the system, followed by the sudden injection of a substantial jailbreak payload utilizing the vast context window limit at a late turn. To prevent historical inertia from suppressing urgent risk surges, production pipelines must integrate an ensemble approach employing Accelerated Failure Time (AFT) models. Utilizing Weibull distributions, AFT models directly shrink the baseline survival time based on the immediate impact magnitude of the most recent turn.

\textbf{Stochastic Auditing for Sub-Threshold Evasion:} The trigger-based cascade architecture conditionally performs the intensive Mahalanobis computation only when the iForest score exceeds the threshold $\alpha$. An adversary adopting a boiling frog micro-perturbation approach could meticulously maintain the anomaly score below $\alpha$ across consecutive turns. Deployments must implement a stochastic checkpoint scheduling mechanism as an asynchronous background process. This mechanism forcefully executes a full covariance precision inspection periodically and completely independent of the iForest score, ensuring cumulative micro-adversarial adjustments are quantified.

\subsection{Standardized Adversarial Simulation Protocol}
To validate the efficacy of the TRIAD framework under unknown zero-day attacks without relying on known attack signatures, we propose a standardized cross-modal context poisoning simulation protocol:

\begin{enumerate}
    \item \textbf{Baseline Parameterization:} Extract normal multimodal interaction embeddings from standard conversational datasets to establish reference covariance structures and baseline isolation dimensions.
    \item \textbf{Orthogonal Drift Injection:} Simulate a stealthy multi-turn attack by gradually injecting an orthogonal noise vector $\delta$ into the visual or textual latent spaces over successive interaction intervals. This bounds individual turn perturbations below standard single-point classification thresholds.
    \item \textbf{Kinematic Profile Evaluation:} Monitor the second-order derivative (trajectory acceleration $a_t$) across the sequence. A valid defense framework must demonstrate that while baseline snapshot detectors fail across extended turns, the continuous accumulation of drift tension reflects a strictly positive acceleration, forcing the survival probability $S(t)$ to exponentially converge to 0 before session completion.
\end{enumerate}

\subsection{Computational Complexity Boundaries}
For real-time agentic AI deployments where latency is a critical constraint, evaluation protocols must verify computational feasibility boundaries. In offline initialization or periodic calibration phases, calculating the robust inverse covariance matrix and constructing the Isolation Forest requires $O(Nd^2 + N\log N)$ time. Crucially, during online real-time inference, evaluating the incoming telemetric vector requires only $O(d^2)$ matrix operations and $O(\log N)$ tree traversal time. This mathematically guarantees deterministic, low-latency execution independent of the length of the semantic interaction history.

\section{Conclusion}

As generative artificial intelligence evolves toward complex, multi-turn Agentic systems, the mechanisms governing alignment and safety must achieve operational parity through advanced, non-linear geometric tracking. Traditional classification models are structurally inadequate against the progressive cross-modal contamination inherent in multi-turn adversarial interactions. By reformulating safety alignment from a reactive classification challenge to a predictive, time-to-failure survival process, the TRIAD framework resolves fundamental blind spots in current architectures. The integration of unsupervised Isolation Forests, Ledoit-Wolf regularized Mahalanobis metrics, and topological kinematics yields an active state-space immune system capable of functioning independent of exhaustive attack signature repositories. Tracking the second-order derivative of semantic divergence provides a robust statistical demarcation between benign creative exploration and continuous adversarial drift, offering a highly efficient mechanism for securing next-generation artificial intelligence.

Despite its theoretical guarantees, practical deployment of the TRIAD framework requires careful handling of operational edge cases, such as manifold heterogeneity, temporal inertia from sudden shock attacks, and sub-threshold evasion techniques. Incorporating structural extensions like Gaussian Mixture Models (GMM) and Accelerated Failure Time (AFT) frameworks helps preserve accuracy during natural conversational domain shifts while neutralizing high-latency adversarial strategies. Transitioning from static telemetric thresholds to dynamic, closed-loop calibration loops ensures resilience without triggering excessive false positives during complex, benign interactions.

Consequently, this methodology marks a decisive shift from static, reactive prompt-filtering to dynamic, predictive survival modeling in AI alignment, establishing the statistical foundations required to operationalize compliance and safety metrics for high-risk autonomous infrastructure. As the industry transitions toward fully autonomous multi-turn systems handling sensitive societal frameworks, defining a mathematical lower bound for adversarial drift detection enables continuous risk auditing in clinical and financial decision-making pipelines, thereby accelerating the safe and responsible integration of generative agents into public infrastructure.

Ultimately, while this paper establishes the mathematical foundation for predictive statistical defense, real-world implementations must evolve into a complex, interdependent mesh network of feedback loops to prevent sophisticated adversaries from reverse-engineering and exploiting the defense mechanism itself.

\bibliographystyle{plainnat}
\bibliography{main}

\end{document}